\documentclass[traditabstract]{aa}
\pdfoutput=1
\usepackage[colorlinks=true,citecolor=blue]{hyperref}
\usepackage{txfonts}
\usepackage{graphicx}
\usepackage{natbib}
\usepackage{epsf}

\def \MJ{M$_{\mathrm{Jup}}$}

\def \msol{M$\mathrm{_\odot}$}

\def \1s{$1\,\sigma$}

\def \t0{T$_0$}

\begin{document}

\title{Eccentricity in planetary systems and the role of binarity \thanks{Based on observations collected with SPHERE on the Very Large Telescope (ESO, Chile). }}
            
\subtitle{Sample definition, initial results, and the system of HD\,211847}

\author{Moutou C.$^{1,2}$, Vigan A.$^2$, Mesa D.$^3$, Desidera S.$^3$, Th\'ebault P.$^4$, A. Zurlo$^{5,6}$, G. Salter$^2$
}

\institute{$^1$ CNRS, CFHT, 65-1238 Mamalahoa Hwy, Kamuela HI 96743, USA\\
$^2$ Aix Marseille Univ, CNRS, LAM, Laboratoire d'Astrophysique de Marseille, Marseille, France\\
$^3$ INAF, Osservatorio Astronomico di Padova, Vicolo dell'Osservatorio 5, 35122 Padova, Italy\\
$^4$ LESIA,  CNRS,  Observatoire  de  Paris,  Universit\'e  Paris  Diderot,
UPMC, 5 place J. Janssen, 92190 Meudon, France\\
$^5$ Nucleo de Astronoma, Facultad de Ingeniera, Univ. D.Portales, Av. Ejercito 441, Santiago, Chile\\
$^6$ Universidad de Chile, Camino del Observatorio, 1515 Santiago, Chile
}

 \date{Received TBC; accepted TBC}
      
\abstract{
 We explore the multiplicity of exoplanet host stars with high-resolution images obtained with VLT/SPHERE. Two different samples of systems were observed: one containing low-eccentricity outer planets, and the other containing high-eccentricity outer planets. 
We find that 10 out of 34 stars in the high-eccentricity systems are members of a binary, while the proportion is 3 out of 27 for circular systems.  Eccentric-exoplanet hosts are, therefore, significantly more likely to have a stellar companion than circular-exoplanet hosts. The median magnitude contrast over the 68 data sets is 11.26 and 9.25, in H and K, respectively, at 0.30 arcsec. The derived detection limits  reveal that binaries with separations of less than 50au are rarer for exoplanet hosts than for field stars. Our results also imply that the majority of high-eccentricity planets are not embedded in multiple stellar systems (24 out of 34), since our detection limits exclude the presence of a stellar companion.   
We detect the low-mass stellar companions of HD\,7449 and HD\,211847, both members of our high-eccentricity sample. HD\,7449B was already detected by \citet{2016ApJ...818..106R} and our independent observation is in agreement with this earlier work. HD\,211847's substellar companion, previously detected by the radial velocity method, is actually a low-mass star seen face-on. The role of stellar multiplicity in shaping planetary systems is confirmed by this work, although it does not appear as the only source of dynamical excitation.}

\authorrunning{Moutou et al.}
\titlerunning{Eccentricity in planetary systems and the role of binarity}
\keywords{ Planetary systems -- binaries: visual -- Stars: imaging -- Techniques: high angular resolution}

\maketitle

\section{Introduction}
The distribution of eccentricities amongst the exoplanet population discovered by the radial-velocity method is intriguing, as nearly circular orbits such as those in the solar system coexist with much more eccentric ones. Approximately 12\% of these RV planets have eccentricities more than 0.5 and 45\% have eccentricities larger than 0.2, which is the maximum eccentricity in the solar system (Mercury). Where planets mostly formed via the core-accretion mechanism in the central part of circumstellar disks, little dynamical excitation is expected \citep{pollack1996}, and simulations of young planetary systems have a clear tendency for coplanarity, circularity, and minimum energy exchange between planets. However, models now exist for more excited systems implying planet-planet scattering \citep{chatterjee2008}, dynamical interactions with a distant stellar companion, and/or the Kozai resonance mechanism \citep{naoz2012,wu2007,kozai1962}. This Kozai mechanism involves the gravitational interaction between a planet and an outer stellar companion that is orbiting at large separations (up to several hundred au) from the central star. The gravitational interactions between both objects induces Kozai oscillations, which gradually pump the planetary orbit to high eccentricity and inclinations while tidal dissipation from the central star circularizes the orbit \citep{wu2007}. For tighter binaries, it is the secular perturbations of the secondary that can force high eccentricities on circumprimary planets (see for ex. HD\,41004 or HD\,196885, \citet{2011CeMDA.111...29T}).

Simulating the dynamical evolution of planetary systems requires the global knowledge of all components of the system and the presence of a distant stellar companion is evidently a significant piece of information improving the accuracy of a model. It is expected from theory that binaries with separations $<$50au should have a strong impact on planet formation around the primary \citep{2014arXiv1406.1357T}. However, observations have revealed that some binaries as tight as 20au indeed harbor planets ($\gamma$ Cephei, HD~196885). Discovering new planet-harboring tight binaries is therefore essential to our understanding of how planet formation unfolds in such extreme conditions. Unfortunately, the stellar multiplicity is generally not known at the time of discovering a new planet with the RV method even though it may seem trivial compared to the planet discovery: while equal-mass short-period binaries are discarded by immediate spectroscopic analysis, long-period stellar companions are more difficult to recognize since their signatures may take tens of years to reveal themselves. For instance, a 0.1 solar mass companion located on a 10$^5$-day orbit from a solar-type star has a semi-amplitude of 60 m/s over 274 years equivalent to non-detectable trend over the typical 2-3 years of RV monitoring. Such secondary components, however, would be located at 0.4" for a 100pc distance system and would have a magnitude difference of six in the H band, making them detectable in direct imaging with instruments such as VLT/SPHERE \citep{beuzit} that provide high-contrast capabilities within one arcsecond. 

Determining any correlation between eccentric systems and the presence of distant massive companions would directly test the formation/migration mechanisms. In addition, peculiar systems with detection of secondary components could be scrutinized with dedicated modeling, as done for $\alpha$ Cen or HD196885 \citep{Thebault2009, Thebault2011}.

Several imaging programs have observed exoplanet-hosting stars in the past. For instance, \citet{Mugrauer2014} and references therein presented SOFI imaging observations of such systems; their observations probed the separations of 30 to 200 arcsec revealing that 13\% of exoplanet host stars in their sample contained distant stellar companions. \citet{narita2012} discovered the stellar companion to HAT-P-7 (with a magnitude difference of approximately six and 3.9" separation) using Subaru IRCS and HiCIAO and were able to revisit the migration history of this system. \citet{2016ApJ...827....8N} focused on the stellar companions of hot-Jupiter systems and found that  wide stellar companions were in excess compared to field stars.
The objective of this work is to assess the presence of any stellar companion in the separation range 0.1--4 arcsec and mass regime $>$ 0.06 solar mass around all targets in two samples, and to statistically compare them. In Section \ref{sample}, we describe the sample selection and strategy. In Section \ref{data}, we present the new data set of VLT/SPHERE high-contrast imaging. In Section 4, we present and discuss candidate detections for all systems with a particular emphasis on HD\,7449 and HD\,211847, which both present short-separation companions. Where no companion is found, we discuss and quantify detection limits. We conclude in Section 5.

\begin{figure}[t]
\begin{center}
\includegraphics[width=\columnwidth]{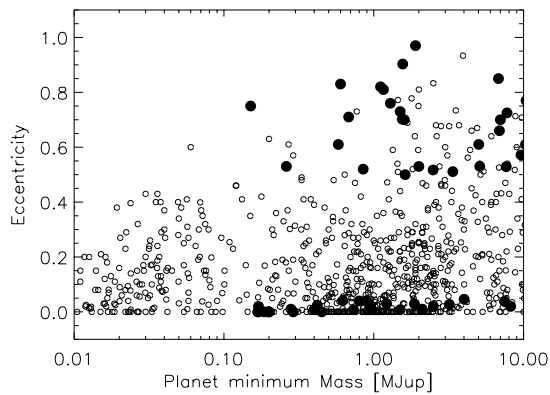}\hfill
\caption{Eccentricity as a function of minimum companion mass of the systems where these parameters are known (open circles), and of the sample of systems considered in this study (filled circles). \label{fig_sample}}
\end{center}
\end{figure}

\section{The sample}
\label{sample}
We have selected all exoplanets in a volume of 100 pc, with host stars brighter than V magnitude of nine, that are able to be observed from the Paranal Observatory (Chile). In this sample we made two groups: planets with very high eccentricity (e$>$0.4, 35 systems) and circular planets (e$<$0.05, 27 systems). We {\it a priori} discarded the systems with intermediate eccentricities, which amounted to 205. When several planets were present in a system (there are 17 multi-planet systems in total) we attribute the system to one sample or the other according to the eccentricity of the outermost planet, since it is the one most susceptible to the effect of massive companions at further distances. We have listed relevant parameters of systems in  Tables~\ref{tab_sample1} and \ref{tab_sample2} and their distribution in the planet mass-eccentricity diagram is shown in Fig.~\ref{fig_sample}. Errors on the parameters are not replicated in these tables but can be found in the Exoplanet Encyclopedia\footnote{http://exoplanet.eu/catalog/}. Note that three additional systems were observed, HD\,47186, HD\,92788, and HD\,168443, which were originally put in the comparison sample, but for which the outermost planet has intermediate eccentricity and does not fulfill any of our criteria anymore (they were misclassified or updated). However, since they were observed their data has been included in the reduction and analysis. Their parameters are given in the bottom lines of Table \ref{tab_sample2}.

The magnitude limit is set to allow the full sample to be observed in a short time as a bad-weather filler program using VLT/SPHERE to either confirm or reject the presence of outer, massive companions, focusing on the closest separation range.

It is interesting to note that the comparison sample contains much more multi-planet systems than the "eccentric sample" (ten vs. four): this is, however, expected due to dynamical interactions \citep[e.g.,][]{2011ApJS..197....8L} and/or detection biases in radial-velocity surveys. A similar observation was made by \citet{desidera2007}.



\begin{table*}
\caption{System parameters for the "eccentric" sample. The selection is based on the eccentricity of the outermost planet. The object with an asterisk $^*$ is actually not a planetary but a stellar companion (see text). There is one line per planet. The age is not always estimated.}
\label{tab_sample1}
\begin{tabular}{lcccccccc}
\hline
Name    &   mH       & Minimum Mp     & $e$    &   Period&         $a$  & Ms   & Dist  &  Age\\
       &              &\MJ   &        &   days&        au  & \msol & pc    &   Gy \\
\hline
\hline
   GJ 3021&    4.99&      3.37&    0.51&    133.71&    0.49&    0.90&   17.62&     8.77\\
   HD 4113&    6.34&      1.56&    0.90&    526.62&    1.28&    0.99&   44.00&     4.80\\
   HD 7449&    6.179&     1.09&    0.80&    1270.&     2.33&    1.05&   39.0&      2.1\\
  HD 20782&   6.18&       1.90&    0.97&    591.9&    1.38&    1.00&   36.02&      7.10\\
$\epsilon$ Eridani&       1.75    & 1.55&    0.70&   2502&    3.39&    0.83&    3.20&   0.66\\
  HD 28254&   6.13&       1.16&    0.81&   1116&    2.15&    1.06&   56.20&        $>$ 3\\
  HD 30562&   4.57&       1.29&    0.76&   1157&    2.30&    1.22&   26.50&       4.00\\
  HD 33283&   6.73&       0.33&    0.48&     18.18&    0.17&    1.24&   86.00&       3.20\\
  HD 39091&   4.42&      10.30&    0.61&   2049&    3.28&    1.10&   18.32&      3.83\\
  HD 43197&   7.33&       0.60&    0.83&    327.8&    0.92&    0.96&   54.90&        $>$ 3\\
  HD 44219&   6.22&       0.58&    0.61&    472.3&    1.19&    1.00&   50.43&        $>$ 3\\
  HD 65216&   6.5&        0.17&    0.02&    152.6&    0.54&    0.92&   34.30&         -\\
  HD 65216&   6.5&        1.26&    0.41&    572.4&    1.3 &    0.92&   34.30&         -\\
  HD 66428&  6.77&        2.82&    0.47&   1973&    3.18&    1.15&   55.00&       5.56\\
  HD 86226&  6.58&        1.50&    0.73&   1534&    2.60&    1.02&   42.48&       -\\
  HD 86264&  6.34&        7.00&    0.70&   1475&    2.86&    1.42&   72.60&      2.24\\
  HD 96167&  6.62&        0.68&    0.71&    498.9&    1.30&    1.31&   84.00&      3.80\\
  HD 98649&  6.49&        6.80&    0.85&    4951    &    5.60&     1.0&   40.30&        2.3\\
 HD 106515A&   5.5&       9.61&    0.57&   3630&    4.59&    0.97&   35.20&        11.7\\
 HD 108147&  5.8&         0.26&    0.53&     10.9&    0.10&    1.19&   38.57&      1.98\\
 HD 129445&  7.24&        1.60&    0.70&   1840&    2.90&    0.99&   67.61&       -\\
 HD 134060&  5.02&        0.035&   0.40&   3.27&    0.044&  $\sim$1.1   &   24.20&        -\\
 HD 134060&  5.02&        0.15&    0.75&   1160.9&    2.23& $\sim$1.1    &   24.20&        -\\
 HD 142022&  6.0&         5.10&    0.53&   1928&    3.03&    0.99&   35.87&     13.30\\
 HD 142415&    5.99&      1.62&    0.50&    386.3&    1.05&    1.09&   34.20&      1.49\\
 HD 148156&  6.49&        0.85&    0.52&   1010&    2.45&    1.22&   53.05&       -\\
 HD 154672&  6.69&        5.02&    0.61&    163.91&    0.60&    1.06&   65.80&       9.28\\
 HD 157172&  6.18&        0.12&    0.46&    104.84&    0.42& $\sim$0.94    &   31.90&        -\\
 HD 181433&  6.18&        0.024&   0.40&   9.37&    0.08&    0.78&   26.15&      -\\
 HD 181433&  6.18&        0.64&    0.28&    962&    1.76&    0.78&   26.15&      -\\
 HD 181433&  6.18&        0.54&    0.48&   2172&    3.00&    0.78&   26.15&      -\\
 HD 187085&  6.18&        0.75&    0.47&    986&    2.05&    1.22&   44.98&      3.30\\
 HD 196067&  5.21&        6.90&    0.66&   3638&    5.02&     1.29&   44.30&         3.3\\
 HD 210277&  4.96&        1.23&    0.47&    442.1&    1.10&    1.09&   21.29&      6.93\\
 HD 211847$^*$&  7.12&  19.20$^*$& 0.69$^*$& 7930$^*$& 7.50$^*$& 0.94& 50.60&   3.00\\
 HD 215497&  6.2&         0.02&    0.16&    3.934&    0.047&    0.87&   44.00&      7.00\\
 HD 215497&  6.2&         0.33&    0.49&    567.94&    1.28&    0.87&   44.00&      7.00\\
 HD 217107&   4.1&        1.33&    0.13&   7.12&    0.073&    1.02&   19.72&     7.32\\
 HD 217107&   4.1&        2.49&    0.52&   4210&    5.27&    1.02&   19.72&     7.32\\
 HD 219077&   4.12&      10.39&    0.77&   5501&    6.22&    1.05&   29.35&        8.90\\
 HD 222582&   6.2&        7.75&    0.73&    572.38&    1.35&    0.99&   42.00&     6.16\\
\hline
\end{tabular}
\end{table*}

\begin{table*}
\caption{System parameters for the comparison sample, and the three intermediate-eccentricity sample stars in the bottom lines. There is one line per planet.}
\label{tab_sample2}
\begin{tabular}{lcccccccc}
\hline
Name    &   mH       & minimum Mp     & $e$    &   Period&         $a$  & Ms   & Distance  &  Age\\
       &              &\MJ   &        &   days&        au  & \msol & pc    &   Gy \\
\hline
\hline
   HD 1461&   5.04&       0.017&    0.00&     5.78&    0.06&    1.08&   23.40&       6.30\\
   HD 1461&   5.04&       0.02&    0.00&     13.51&    0.11&    1.08&   23.40&       6.30\\
   HD 4208&   6.24&       0.80&    0.04&    829&    1.70&    0.87&   33.90&      4.47\\
  HD 10180&   5.93&       0.04&    0.045&   5.76&    0.064&    1.06&   39.40&      4.30\\
  HD 10180&   5.93&       0.04&    0.088&   16.35&   0.128&    1.06&   39.40&      4.30\\
  HD 10180&   5.93&       0.08&    0.026&   49.7&    0.270&    1.06&   39.40&      4.30\\
  HD 10180&   5.93&       0.08&    0.135&   122.8&   0.493&    1.06&   39.40&      4.30\\
  HD 10180&   5.93&       0.07&    0.19&   601.2&    1.42&    1.06&   39.40&      4.30\\
  HD 10180&   5.93&       0.20&    0.08&   2222&    3.40&    1.06&   39.40&      4.30\\
  HD 11964&   4.5&        0.079&    0.3&   37.91&    0.23&    1.12&   33.98&      9.56\\
  HD 11964&   4.5&        0.62&    0.04&   1945&    3.16&    1.12&   33.98&      9.56\\
  HD 20794&   2.6&        0.009&    0.00&     18.31&    0.12&    0.85&    6.06&      14.00\\
  HD 20794&   2.6&        0.008&    0.00&     40.1&    0.20&    0.85&    6.06&      14.00\\
  HD 20794&   2.6&        0.015&    0.00&     90.31&    0.35&    0.85&    6.06&      14.00\\
  HD 23079&  5.81&        2.50&    0.02&    626&    1.50&    1.10&   34.80&     6.53\\
  HD 38801&  6.37&       10.70&    0.00&    696.3&    1.70&    1.36&   99.40&      4.67\\
 BD-061339&  6.52&        0.027&    0.00&    3.873&    0.043&    0.70&   20.00&       4.40\\
 BD-061339&  6.52&        0.17&    0.00&    125.94&    0.435&    0.70&   20.00&       4.40\\
  HD 60532&  3.3&        9.21&    0.278&    201.8&    0.77&    1.44&   25.70&      2.70\\
  HD 60532&  3.3&        21.8&    0.04&    607.06&    1.58&    1.44&   25.70&      2.70\\
  HD 73256&   6.37&      1.87&    0.03&      2.55&    0.04&    1.05&   36.50&       0.83\\
  HD 75289&  5.13&       0.42&    0.02&      3.51&    0.05&    1.05&   28.94&        4.96\\
  HD 76700&  6.37&       0.20&    0.00&      3.97&    0.05&    1.00&   59.70&      4.51\\
  HD 82943&  5.18&       14.4&    0.42&   219.3&    0.746&    1.18&   27.46&        3.08\\
  HD 82943&  5.18&       14&    0.203&   442.4&    1.19&    1.18&   27.46&        3.08\\
  HD 82943&  5.18&       0.29&    0.00&   1078&    2.14&    1.18&   27.46&        3.08\\
  HD 83443&  6.58&       0.40&    0.01&      2.99&    0.04&    0.90&   43.54&       2.94\\
  HD 85390&   6.61&      0.13&    0.41&   788&    1.52&    0.76&   33.96&         7.20\\
  HD 85390&   6.61&      0.20&    0.00&   3700&    4.23&    0.76&   33.96&         7.20\\
  HD 86081& 7.42&        1.50&    0.01&      2.14&    0.04&    1.21&   91.00&     6.21\\
 HD 104067&  5.75&       0.19&    0.00&     55.81&    0.26&    0.79&   20.80&       4.33\\
 HD 109749&  6.57&       0.28&    0.01&      5.24&    0.06&    1.20&   59.00&      10.30\\
 HD 117618&  5.82&       0.18&    0.42&    25.8&    0.176&    1.05&   38.00&        3.88\\
 HD 117618&  5.82&       0.20&    0.00&    318&    0.93&    1.05&   38.00&        3.88\\
 HD 121504&  6.03&       1.22&    0.03&     63.33&    0.33&    1.18&   44.37&      1.62\\
 HD 150433&  5.72&       0.17&    0.00&   1096.2&    1.93&     $\sim$0.98&   29.60&       -\\
 HD 159868&  5.57&       0.73&    0.15&    352.3&    1.0&    1.09&   52.70&     8.10\\
 HD 159868&  5.57&       2.10&    0.01&   1178.4&    2.25&    1.09&   52.70&     8.10\\
 HD 179949&   5.1&       0.95&    0.02&      3.09&    0.05&    1.28&   27.00&       2.05\\
 HD 192263&   6.5&       0.73&    0.01&     24.36&    0.15&    0.81&   19.90&       0.57\\
HIP 105854& 3.11&        8.20&    0.02&    184.2&    0.81&    2.10&   80.84&         -\\
 HD 212301& 6.76&        0.45&    0.00&      2.25&    0.04&    1.05&   52.70&        5.90\\
  91 Aqr &    1.9&       3.20&    0.03&    181.4&    0.70&    1.40&   45.90&       3.56\\
\hline
  HD 47186&  6.1&        0.07&    0.04&    4.08&    0.05&    0.99&   37.84&      -\\
  HD 47186&  6.1&        0.35&    0.25&   1353.6&    2.39&    0.99&   37.84&      -\\
  HD 92788& 6.01&        0.90&    0.04&    162&    0.60&    1.13&   32.82&        3.78\\
  HD 92788& 6.01&        27.7&    0.33&    325.8&    0.97&    1.13&   32.82&        3.78\\
 HD 168443&  5.32&        7.66&    0.53&     58.11&    0.29&    1.00&   37.38&        9.80\\
 HD 168443&  5.32&        29.46&   0.21&     1749&    2.83&    1.00&   37.38&        9.80\\
\hline
\end{tabular}
\end{table*}

A literature search reveals that several of these planet hosts are in multiple stellar systems. The compilation of known binary (and one triple) systems is given in Table \ref{tab_known_binaries}. Their projected distance ranges from 21 to 9100au, with a median of $\sim$600au. There are ten known multiple systems in the eccentric sample, and three in the comparison sample. Seven of these systems have $\Delta$K less than 2.0.

\begin{table*}
\begin{center}
\caption{Known visual binaries. The top lines corresponds to stars in the `eccentric' sample while those in the bottom lines are from the comparison sample. References are  \citet{2006A&A...456.1165C} (Ch06), \citet{2014MNRAS.439.1063M} (M14), \citet{2016ApJ...818..106R} (R16), \citet{2014A&A...569A.120L} (L14), \citet{2007MNRAS.378.1328M} (M07), \citet{2012A&A...546A.108D} (D12), \citet{2006A&A...447.1159E} (E06), \citet{2006ApJ...646..523R} (R06), \citet{2012A&A...542A..92R} (R12), and \citet{2004A&A...425..249M} (M04), or the Henry Draper Catalog. }
\label{tab_known_binaries}
\begin{tabular}{lcccc}
\hline
Name      & Separation & Proj. Distance &$\Delta$K &Reference \\
          &   arcsec   &   au           & mag      &          \\
\hline
GJ 3021   &  3.86    & 68               & 5.0     & Ch06  \\
HD 4113   & 49.0     & 2156             & 2.25    & M14 \\
HD 7449   &   0.545  & 21.2             & 1.886   & R16 \\
HD 20782  & 253      & 9108             & 0.726   & HD catalog  \\
HD 28254  &   4.3    & 242              & 0.814   & L14 \\
HD 65216  &   8.2    & 281              & 6.31    & M07  (triple)\\
HD 106515 &  6.89    & 243              & 0.116   & D12 \\
HD 142022 &  20.16   & 723              & 1.496   & E06 \\
HD 196067 &  16.8    & 744              & 0.63    & HD catalog \\
HD 222582 & 109.45   & 4597             & 3.41    & R06 \\
\hline
HD 11964  &  29.4    & 999              & 3.107    & R12 \\
HD 75289  &  21.54   & 623              & 5.819    & M04 \\
HD 109749 &   8.35   & 493              & 1.445    & L14 \\
\hline
\end{tabular}
\end{center}
\end{table*}

\section{Observations and analysis}
\label{data}
\subsection{Instrument and strategy}

We used the adaptive-optics instrument SPHERE at the Very Large Telescope (ESO, Chile) \citep{beuzit} in the ESO programs 95.C-0476 and 96.C-0249. SPHERE was used in the \texttt{IRDIFS-EXT} mode. It uses the two near-infrared focal instruments IRDIS and IFS, the former in the dual-band imaging (DBI) in K-band \citep{vigan2010} and the later covering the YJH spectral bands \citep{claudi2008}. The filters used in Differential Band Imaging with IRDIS in this mode are K1 (2.11 $\mu$m) and K2 (2.25 $\mu$m), with 0.1 $\mu$m bandpasses.

All observations were performed in pupil-stabilized mode with the apodized pupil Lyot coronograph N-ALC-YJH-S \citep{soummer2005}, which uses a coronagraphic mask of diameter 185 mas. 
The IRDIS detector was dithered on a 4$\times$4 pattern to reduce the effect of the residual flat-field noise. 

The observation sequence consisted of: \begin{itemize}
    \item a coronographic image sequence on the source of typically 20 min ;
    \item an out-of-mask flux reference (PSF image), where the field is offset so that the primary star is visible with neutral densities inserted as needed, to allow a photometric reference;
    \item a centering image, where four symmetric
satellite spots are created by introducing a periodic modulation on the deformable mirror. This data is used to determine the position of the star center behind the coronagraph and the center of field rotation;
    \item another coronographic image sequence;
    \item a sky observation on a nearby region, mostly useful for the K band observations with IRDIS.
\end{itemize}

For each step of this sequence, we optimized the observing parameters of both IRDIS and IFS (detector integration time, number of iterations, dithering pattern, and neutral densities) so that no part of the instrument is idle for too long despite their slightly different sensitivities. 

Looking for relatively low contrast images, we set loose constraints for observations: seeing up to 1.4", variable extinction, airmass up to 2.0 and no timing constraint on the field rotation velocity (as appropriate for a bad-weather program). The seeing and field rotation conditions at which observations were collected are listed in Tables \ref{log} and \ref{logctd}. Seeing conditions were sometimes extreme ($>$1.5") during these observations, during which the AO loop broke open for parts of some sequences. 

The full sample presented in section \ref{sample} has been observed between April 2015 and March 2016 in service mode.  
We used calibration data collected in the morning (darks, instrumental background, and flat-field images) in the standard way.

\subsection{Data reduction}
\subsubsection{IRDIS}
\label{IRDISdata}
We used the SPHERE DRH pipeline \citep{pavlov2008} to reduce the calibration frames, then the LAM-ADI pipeline described in \citet{vigan2012} for the IRDIS science images. Each of the images in the coronagraphic observing sequences were  sky-background subtracted  and  divided  by  the  master K1-K2 flat-field.  Bad pixels were detected  using
bad pixel maps created with the DRH and corrected by replacing them with
the  median  of  neighboring  good  pixels. All processed images were then aligned to a common center using the star center data and photometrically calibrated using the PSF observation. At this point, it was necessary to qualify each frame of the data cubes individually and to reject images where the loop was open or the extinction too large. Frame selection is done separately on IRDIS and IFS, since there is no fine synchronization between the acquisition of images. Details on the criteria used for IRDIS frame selection is given in Appendix \ref{appA}.

Because of the loose constraint on the observation time with respect to meridian passage and the short exposure time, the amount of field of view rotation in each sequence was usually small ($<$ 10 degrees in more than half of the data set). As a consequence, we did not perform a sophisticated data analysis based on angular differential imaging (ADI, \citet{marois2006}). The images were derotated according to their individual parallactic angle, median combined, and then a spatial filter was applied in a box of size 5 lambda/D to remove large spatial scale structures.

\subsubsection{IFS}

IFS data were reduced using the DRH pipeline \citep{pavlov2008} for all the standard calibrations: master dark, master detector 
flat definition of the position of each spectra, wavelength calibration for each spectra and creation of the instrumental flat field. 
From this procedure we obtained a calibrated data cube composed of 39 monochromatic images for each frame. The calibrated 
datacubes were then properly registered exploiting the images with satellite spots following the method exposed in \citet{Mesa2015}.  
The speckle noise was subtracted applying the principal components analysis \citep[PCA-][]{Soummer2012} algorithm adapted to the IFS 
case as devised in \citet{Mesa2015} to be able to apply both ADI and spectral differential imaging 
\citep[SDI-][]{racine1999} at the same time. For the target with reduced datasets (e.g., less than 20 frames) we decided not to perform any frame 
selection. For larger datasets composed of some tens of frames we preferred to bin different images in such a way that the rotation
angle between two different resulting frames was of the order of $\sim$0.5$^{\circ}$-0.6$^{\circ}$. 


\section{Candidate detection}

We detected two candidate companions at short (sub-arcsec) separation around H\,211847 and HD\,7449, in addition to 49 sources at wider separations (up to 6").

\subsection{Small-separation companion candidates}
\subsubsection{HD 211847}

\begin{table*}
\caption{Astrometric values for HD\,211847\,B using IRDIS and IFS. Errors of 5mas and 0.5 degrees are estimated for the distances and position angles, respectively.}
\label{tab_astrohd211847}
\begin{tabular}{c c c c c}
\hline
    &  $\Delta\alpha$ (mas) &   $\Delta$Dec (mas) &  Separation (mas)     & Position Angle ($^{\circ}$)  \\
\hline
\hline
 IRDIS (K1)  &  -53  & -213   & 219   &  193.0 \\
 IRDIS (K2)  &  -49  & -213   & 218   &  193.0 \\
 IFS         &  -53  & -216   & 222   &  193.8 \\ 
\hline
\end{tabular}
\end{table*}

\begin{table*}
\caption{Photometric values for HD\,211847\,B using IRDIS and IFS and corresponding values in mass. }
\label{tab_photohd211847}
\begin{tabular}{c c c c c c}
\hline
                     &  Y      &        J   &  H          &   K1   &  K2  \\
\hline
\hline                     
Absolute magnitude   &  9.87   &  9.29  &  8.93   &  8.43  & 8.24 \\
Mass (\MJ)     & 150.4   &  155.5 &  142.4  &  162.7 & 164.0\\
\hline
\end{tabular}
\end{table*}

\begin{figure}
\includegraphics[width=\columnwidth]{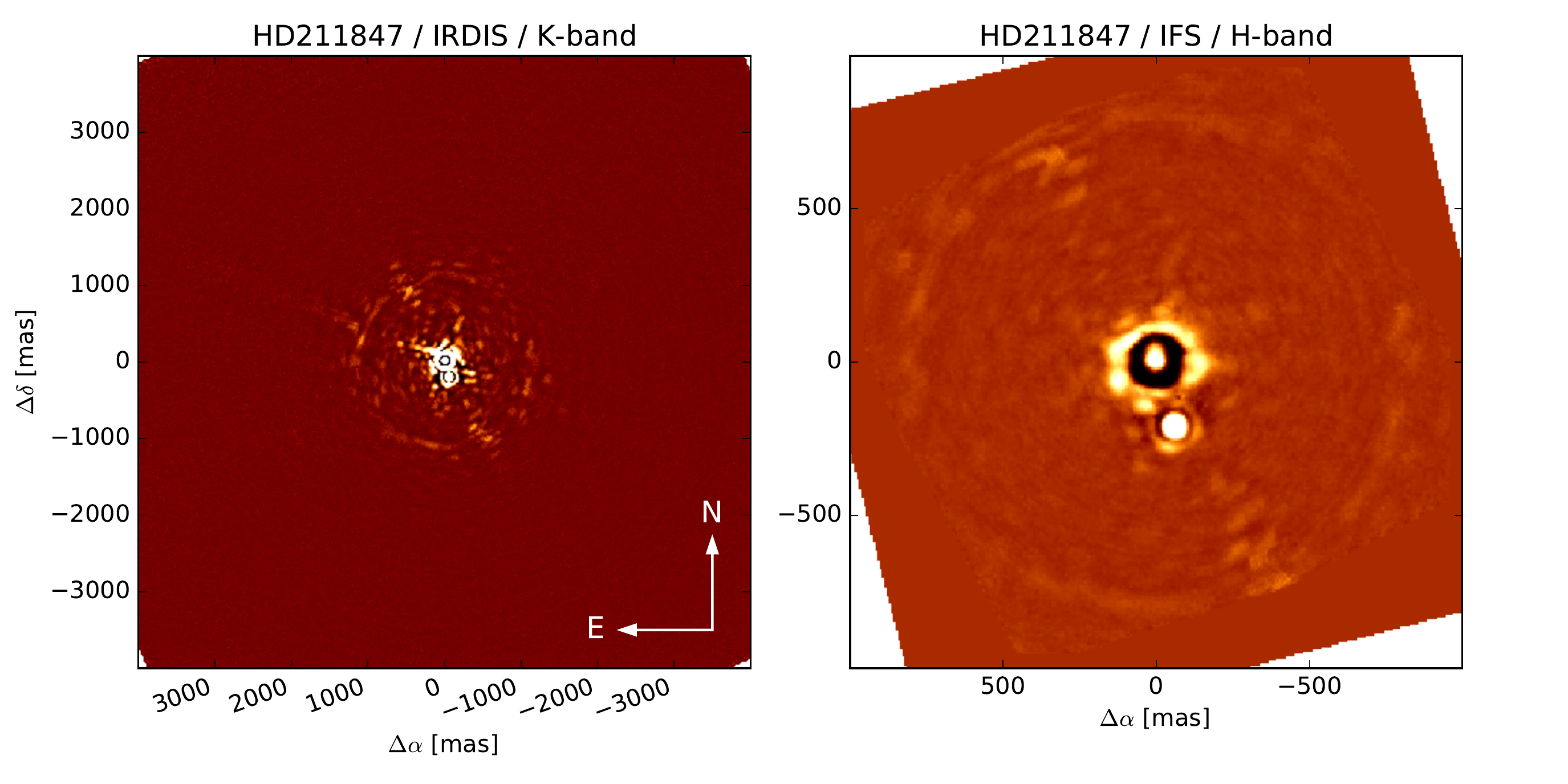}
\caption{IRDIS imaged processed by Angular Differential Imaging (left, field of view 11 arcsec) and IFS H reduced image (right, field of view 1.7 arcsec) of HD211847. North is up and East is left.}
\label{hd211847}
\end{figure}

A candidate companion is detected in the image of HD\,211847 with both IRDIS and IFS, as shown in Figure \ref{hd211847}. The angular separation is 219.6$\pm$2 mas and magnitude difference ranges from 4 to 5. Its astrometric values with respect to HD\,211487 are reported in Table~\ref{tab_astrohd211847}. 
The separation corresponds to a projected separation of 11.3au.

In the first row of Table~\ref{tab_photohd211847}, we report the absolute magnitudes derived for the companion source of HD\,211847 in five different spectral bands, assuming a common distance modulus. Values for Y, J, and H bands are derived from IFS data making an average on the contrast obtained for all of the spectral channels in the wavelength range corresponding to that band. Values for K1 and K2 bands are derived from IRDIS data. Using the BT-Settl models \citep{2014IAUS..299..271A} and assuming an age of 3~Gyr (see Table~\ref{tab_sample1}), we were able to derive the mass of the companion. We find a companion mass of 155 $\pm$9 \MJ\ or 0.148$\pm$0.008 \msol. Masses derived for each infrared band are reported in Table~\ref{tab_photohd211847}.

\begin{figure}
\includegraphics[width=\columnwidth]{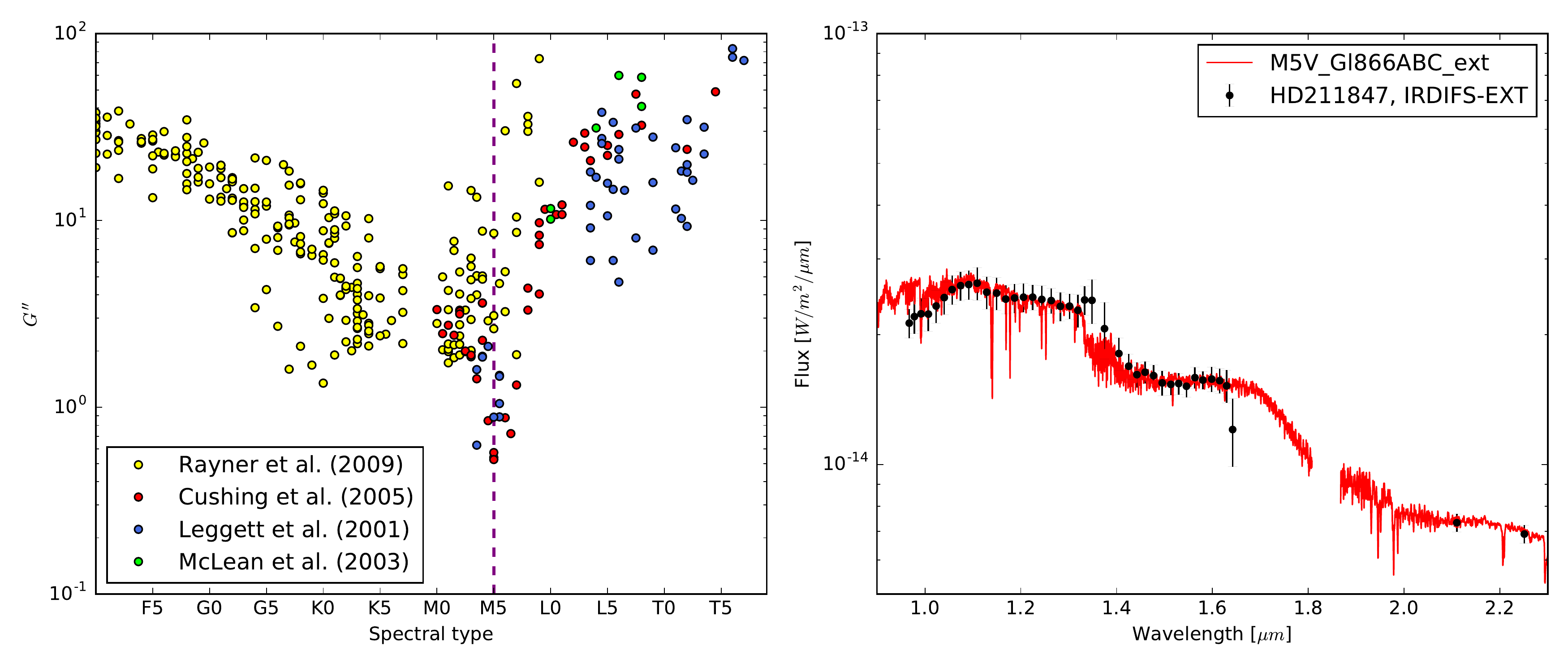}
\caption{(Left) Goodness-of-fit ($G''$ factor, see text for details) with respect to the spectral type for the comparison of our IRDIFS data of the HD\,211847 candidate to spectral templates from various stellar and brown-dwarfs libraries; the best fit is shown by a vertical line. (Right) The best-fit spectrum of Gl\,866 (plain red line) compared to the extracted data points (black dots).}
\label{hd211847spectrum}
\end{figure}

We derived a spectrum for the companion by joining the 39 spectral channels of the IFS to the two from IRDIS. We performed a spectral classification by comparing this spectrum to a sample of template spectra of field stars and brown dwarfs from the IRTF stellar library \citep{cushing2005,rayner2009}, NIRSPEC brown dwarf spectroscopic survey \citep{mclean2003}, and from \citet{leggett2001}. Following the procedure described in \citet{vigan2016}, we use the goodness-of-fit $G''$ indicator defined by \citet{bowler2010} to compare our IRDIFS data to the templates. This indicator enables us to compare SEDs with an inhomogeneous wavelength sampling and with measurement errors on both the templates and the object. The result of this procedure is plotted in Fig.~\ref{hd211847spectrum}, with the $G''$ distribution as a function of spectral type (left panel) and the comparison of the best fit with our data points (right panel). There is a clear minimum in the $G''$ distribution in the M3--M6 range, with a best fit to Gl~866, an M5V star from the \citet{rayner2009} library. This best fit is in good agreement with the mass derived above\footnote{\url{http://www.pas.rochester.edu/~emamajek/EEM_dwarf_UBVIJHK_colors_Teff.txt}} from photometry.

In the radial-velocity discovery paper \citep{2011A&A...525A..95S}, the best-fit solution has a semi-major axis of 7.5$\pm$1.5au but other solutions between 4 and 50au cannot be excluded due to the incomplete time coverage of the orbit. The minimum mass of the companion was then found to be 19.2 \MJ\ (best solution), with corresponding minimum and maximum values of 17 and 23 \MJ.

From the results of our high-contrast imaging study, we find that the radial-velocity companion detected by \citet{2011A&A...525A..95S} is compatible with the one seen in SPHERE images and so is actually a low-mass star (M5 spectral type) instead of a substellar companion. It implies an angle of 7$^{\circ}$ for the inclination of the binary orbit, and therefore an almost face-on system. With an updated radial-velocity data set and the imaging parameters, it would be possible to better constrain the stellar companion. However, assuming it is not a planet host system anymore, there would be little interest to go further.\\

In the light of this new result, and having identified the cause of the main radial-velocity signal detected on HD\,211847 as being of stellar origin, we went back to the CORALIE and HARPS time series in order to search for upper limits on other planetary candidates. We used the DACE analysis tools \citep{2016A&A...590A.134D,2016A&A...585A.134D} to revisit the signal. A one-Keplerian model fit gives a period for the long-term companion of 3134$\pm$566 days and a reduced $\chi^2$ of 5.93. There is some noise in the residuals 
that was interpreted as stellar noise in \citet{2011A&A...525A..95S}. In the periodogram of residuals there are two main peaks at 18.5 and 2.016 days. The first one could be related to the rotation period of the star. We then explored the 2.016d signal further. By adding a second short-term Keplerian in the fit, we find a solution whose reduced $\chi^2$ is 2.01, a significant improvement from the one-Keplerian model, even though the obtained best fit would give an unprobable eccentric orbit. The star being active and activity signals not available, we cannot come to a conclusion on the nature of the 2.016d signal and leave it for future work.

The residual radial-velocity time series after the long-term signal is removed can be translated into a detection limit for another giant planet in the system. With 32 measurements spread over a time span of 2635 days and a residual scatter of 11.5 m/s, a 0.8 \MJ\, planet in a 4au orbit would have been at the detection limit.

\begin{figure}
\includegraphics[width=\columnwidth]{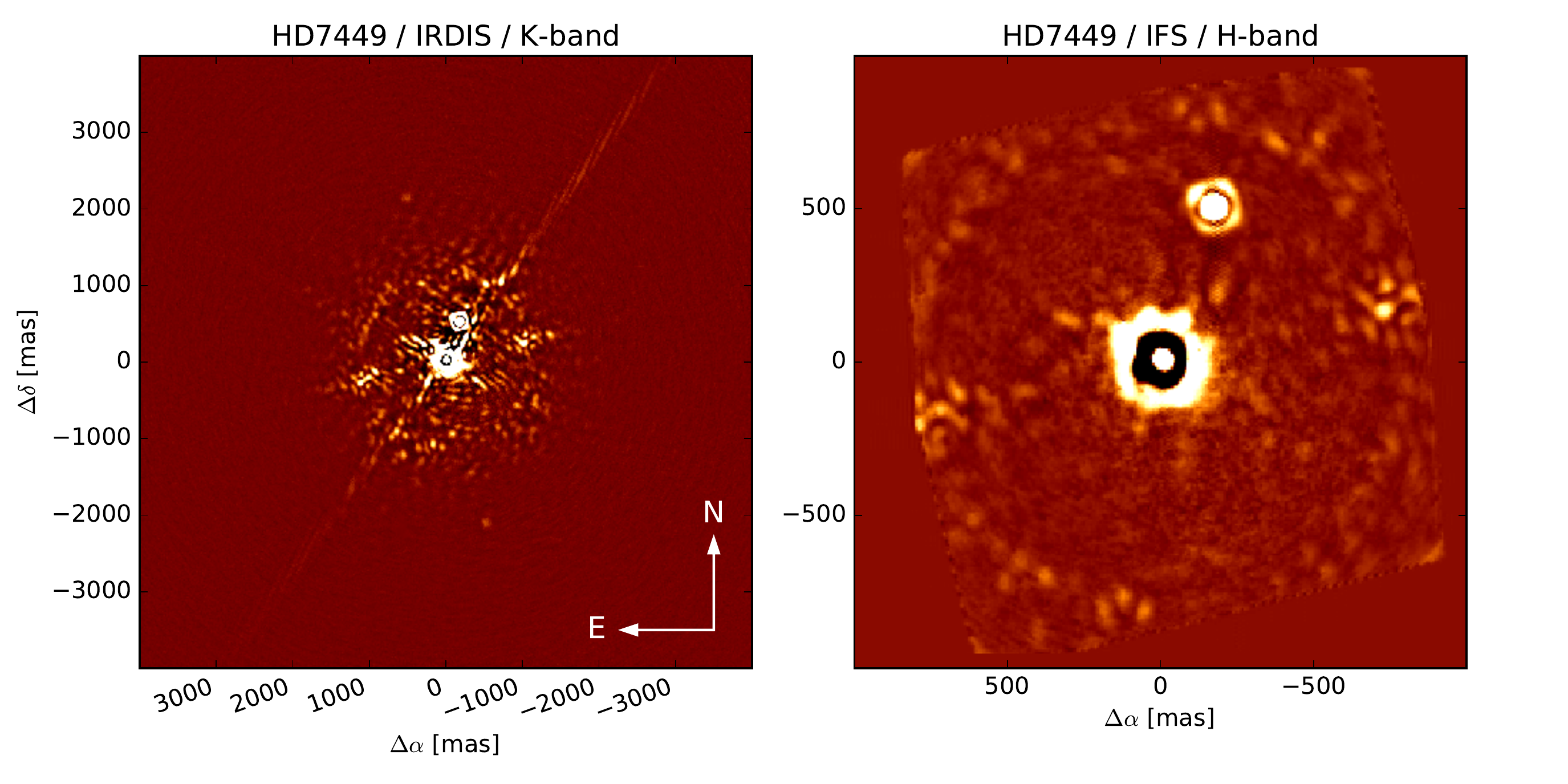}
\caption{IRDIS image processed by Angular Differential Imaging (left, field of view 11 arcsec) and IFS H image (right, field of view 1.7 arcsec) of HD7449.}
\label{hd7449}
\end{figure}


\begin{figure}
\includegraphics[width=\columnwidth]{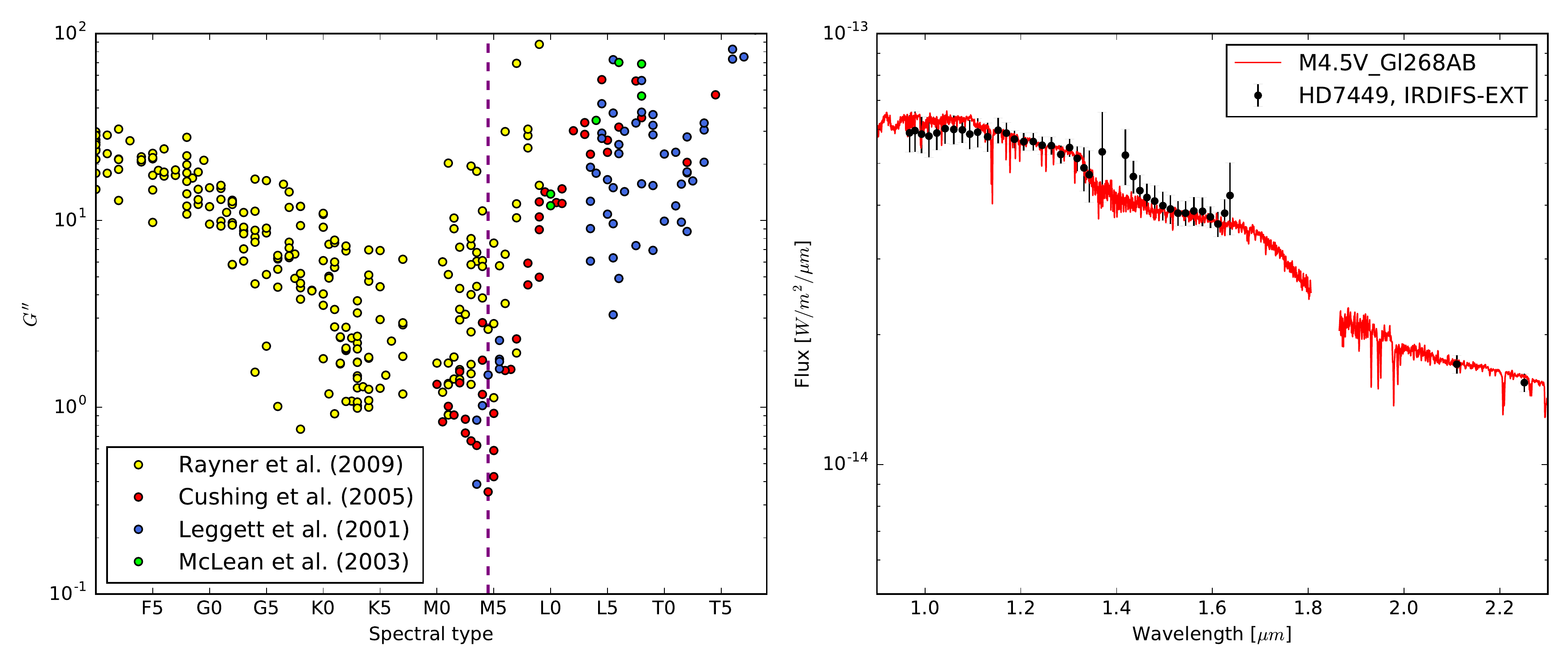}
\caption{(Left) Goodness-of-fit ($G''$ factor, see text for details) with respect to the spectral type for the comparison of our IRDIFS data of the HD\,7449 candidate to spectral templates from various stellar and brown-dwarf libraries; the best fit is shown by a vertical line. (Right) The best-fit spectrum of Gl\,268 (plain red line) compared to the extracted data points (black dots).
}
\label{hd7449spectrum}
\end{figure}

\subsubsection{HD7449}

\begin{table*}
\caption{Astrometric values for HD\,7449\,B using IRDIS and IFS.}
\label{tab_astrohd7449}
\begin{tabular}{c c c c c}
\hline
    &  $\Delta\alpha$ (mas) &   $\Delta$Dec (mas) &  Separation (mas)     & Position Angle ($^{\circ}$)  \\
\hline
\hline
 IRDIS (K1)  &  -171  & 505   & 533   &  341.3 \\
 IRDIS (K2)  &  -170  & 508   & 536   &  341.5 \\
 IFS         &  -163  & 499   & 525   &  341.9 \\ 
\hline
\end{tabular}
\end{table*}

\begin{table*}
\caption{Photometric values for HD\,7449\,B using IRDIS and IFS and corresponding values in mass.}
\label{tab_photohd7449}
\begin{tabular}{c c c c c c}
\hline
                     &  Y      &        J   &  H          &   K1   &  K2  \\
\hline
\hline                     
Absolute magnitude   &  9.49   &  9.02  &  8.43   &  8.10  & 7.95 \\
Mass (\MJ)           & 176.7   &  176.0 &  177.8  &  188.7 & 187.7\\
\hline
\end{tabular}
\end{table*}

A candidate companion is detected both with IRDIS and IFS, as shown in Figure \ref{hd7449}. It lies at 531$\pm$6 mas from the primary, with a magnitude contrast of 5.

Its astrometric values with respect to HD\,7449 are reported in Table~\ref{tab_astrohd7449}. The angular separation corresponds to a projected distance of 20.4au. In Table~\ref{tab_photohd7449}, we report the absolute magnitudes derived for HD\,7449\,B in five different spectral bands. Values for band Y, J, and H are derived from IFS data making a median on the contrast obtained for all the spectral channels in the wavelength range corresponding to that band. Values for K1 and K2 bands are derived from IRDIS data. Using the BT-Settl models and assuming an age of 2 Gyr, we were able to derive the mass of the companions, which are reported in the second row of Table~\ref{tab_photohd7449}.

Our values for HD\,7449\,B are in good agreement with those reported in \citet{2016ApJ...818..106R}. The relative position in right ascension and declination also agrees with the expected proper motion of the primary (Fig. 3 in \citet{2016ApJ...818..106R}).

We thus infer that the outer radial-velocity companion detected by \citet{2011A&A...535A..55D} around HD\,7449 and later re-analyzed in \citet{2016ApJ...818..106R} is seen in SPHERE images and is a low-mass star with a mass of $0.173\pm0.006$~\msol. The inner planet HD\,7449\,Ab has a minimum mass of 1.09$_{-0.19}^{+0.52}$ \MJ\, and period of 1200$_{-12}^{+16}$ days, eccentricity of 0.8 and semi-major axis of 2.33$_{-0.02}^{+0.01}$ \citep{2016ApJ...818..106R}. If the measured angular separation was the semi-major axis of HD\,7449\,B, the ratio of semi-major axis would be approximately 9. However, the orbit of the stellar companion is not yet constrained and this ratio could be anywhere in the range 2-20.

As for HD\,211847\,B, we derived a spectrum of HD\,7449\,B using both IRDIS and IFS data and compared it to the same template spectra. The final result from this procedure is shown in Figure~\ref{hd7449spectrum} where the HD\,7449\,B spectrum is compared
to the best fit spectrum from the adopted sample, the M4.5 star Gl268. 

\subsection{Wide-separation sources}

With a larger field of view, IRDIS images contain more detected sources than IFS. For the sake of completeness, we have identified and measured all sources definitely detected in the IRDIS images. 
The list of these sources is given in Table \ref{ccirdis}. The angular separation measured for these sources has been translated into physical distance in the case where the source would be at the distance of the primary. This is most likely not the case but allows for identification of possible candidates of interest for dynamical studies, stellar companions at less than $\sim$50au being more susceptible to having a dynamical impact. Apart from HD\,211847 and HD\,7449, there are no further systems at less than 50au.

The probability of contamination by distant background stars was evaluated with the Besancon Galaxy Model \citep{2003A&A...409..523R} and Trilegal \citep{2016AN....337..871G}. Even for the most crowded fields (around HD\,108147 or HD\,148156), the star counts in the typical distance and magnitude range of these sources is of the order of 1.5 to 3\%. With such a narrow range of probabilities, these stellar counts are thus not very useful for identifying whether such sources are likely due to local environment of the primary target or distant contaminants. Checking the galactic coordinates, we found that almost all targets with companion candidates had galactic latitudes in the range [-20$^\circ$,+20$^\circ$] and their fields at wider angles are also very crowded. There are five exceptions: GJ 3021, HD 28254, HD 7449, HD 211847, and HD 96167. The companion candidates of the first four systems are actually gravitationally bound secondary stars, identified either from previous work (see Table 3) or from this work (see Section 4.1). HD 96167 has a galactic latitude of 44$^\circ$ and its wide field is not crowded. If bound, its companion would have an absolute magnitude of 20.3 in the K band, which corresponds to a 30\MJ\ object at an age of 3.8 Gy \citep{2014IAUS..299..271A}. Follow-up observations could confirm whether this source is a wide brown dwarf companion of HD 96167 or a background star. In the following, we do not consider it as a companion of relevance for binarity frequency. 

In addition, we checked the absolute magnitude and corresponding mass of all companion candidates in Table \ref{ccirdis}, where these sources are at the same distance as the primary star, and found no mass in the range of stellar companions. When no age had been estimated, we assumed an arbitrary age of 4 Gy. For all except the companions flagged as "confirmed" in Table \ref{ccirdis}, the detected sources are considered background sources.

\begin{table*}
\caption{Companion candidates found in IRDIS images: position angle, separation from the primary star, corresponding distance in au (in those cases where the source is at the distance of the primary target from the Sun), and magnitude difference in the K band. The top lines correspond to the eccentric sample, the middle lines to the comparison sample, and the last line to the intermediate. }
\begin{center}
\label{ccirdis}
\begin{tabular}{lccccl}
\hline
Star    &   Pos. Ang.&  Separation & Distance & $\Delta$K & Comment\\
        & degrees    &  arcsec     & au       &    &\\
\hline
GJ3021      &  353  &   3.97  &      110.7   &    5.8 & confirmed, see Table 1\\
HD7449      &  325  &   0.25  &       9.62   &    4.90    & confirmed, see Table 8   \\
HD28254     &  260  &   4.89  &      274.7  & $>$ 4.0 & confirmed, see Table 1\\
HD96167   &    308  &   3.43  &    289.6    &   11.70 & to be followed-up \\
HD108147  &    3    &   4.90  &     189.0   &     10.90 &likely background\\
HD108147  &    31   &   3.59  &     138.5    &     12.60&likely background\\
HD108147  &    147  &   4.84  &     186.7       &     12.20&likely background\\
HD108147  &    289  &   2.75  &    106.3 &     11.85&likely background\\
HD108147  &    305  &   3.46  &    133.6 &     12.70&likely background\\
HD108147  &     92  &   3.12  &    120.9 &     12.10&likely background\\
HD108147  &     41  &   3.12  &    120.7 &     11.40&likely background\\
HD129445    &  318  &   3.25  &      220.1 &     9.25&likely background\\HD129445    &  185  &   3.48  &      237.6 &     10.7&likely background\\HD129445    &  247  &   1.93  &      131.3 &    11.00&likely background\\HD134060  &    204  &   4.25  &    103.2 &      10.80&likely background\\HD134060  &    278  &   2.58  &     62.5 &      11.80&likely background\\HD134060  &    288  &   3.48  &     84.6 &      11.60&likely background\\HD142415    &  329  &   1.58  &      54.0 &    10.80&likely background \\HD142415    &  317  &   4.76  &    161.9 &    9.95&likely background \\
HD148156   &   121  &   3.25  &    173.0 &     9.40&likely background\\
HD148156    &  123  &   2.56  &    135.8 &     9.70&likely background \\
HD148156    &  245  &   5.06  &    269.2 &     9.90&likely background\\
HD148156    &  113  &   2.41  &    128.0 &     10.60&likely background\\
HD148156    &  146  &   3.81  &    202.1 &    10.95&likely background     \\
HD148156    &  155  &   1.94  &    102.9 &    11.30 &likely background   \\
HD148156    &  7    &   2.85  &    151.2 &    11.30 &likely background\\
HD148156    &  345  &   2.45  &    130.0   &    11.00&likely background \\
HD148156    &  330  &   2.47  &    131.0 &    11.20 &likely background\\
HD148156    &  320  &   3.57  &    189.4 &    11.50 &likely background\\
HD148156    &  250  &   2.10  &    111.4 &    11.40 &likely background\\
HD154672    &  191  &   4.57  &    302.3 &     10.40&likely background\\
HD157172    &  14   &   2.82  &       89.9 &    10.10&likely background\\HD157172    &  241  &   4.26  &      136.3 &    12.80&likely background\\HD157172    &  177  &   5.02  &      160.1 &    10.95&likely background\\HD211847    &  194  &   0.22  &     11.2 &      4.80   & confirmed, see Table 6\\
\hline
HD60532     & 181  &    3.60  &       92.6  &    13.50&likely background\\
HD73256     & 48   &   6.17  &        225.2 &    9.50 &likely background\\
HD73256     & 311   &   6.09  &       222.  &   11.80 &likely background\\
HD76700     & 209   &   4.96  &      296.7  &    9.20&likely background\\HD85390     & 352   &   4.77  &     162.1  &    10.75&likely background\\HD85390     & 37   &   5.41  &      184.2  &   12.50&likely background\\
HD121504    & 194  &   4.50  &      199.7 &    6.50 &likely background\\
HD121504    & 350  &  3.94    &     174.8 &   10.95  &likely background\\HD121504    & 80  &   3.15    &     139.8 &   11.35  &likely background\\HD159868    &  328  &   4.77  &     251.9 &     12.20&likely background\\HD159868    &  222  &   3.62  &     390.8  &     12.90&likely background\\
HD159868    &  145  &   5.93  &      312.5 &     11.75&likely background\\
HD159868    &  128  &   6.52  &      343.6 &     10.90&likely background\\
\hline
HD168443    &  95  &    2.24  &     83.7 &   11.4  &likely background\\
HD168443    &  214  &    2.11  &     78.9 &     11.95&likely background\\HD168443    &  269  &    1.15  &     42.9 &     9.80 &likely background\\\hline
\end{tabular}
\end{center}
\end{table*}

\subsection{Contrast plots}

Finally, we present here all detection plots obtained with both instruments, and translate them into detection limits for stellar, or substellar companions. 



\begin{figure}[t]
\begin{center}
\includegraphics[width=\columnwidth]{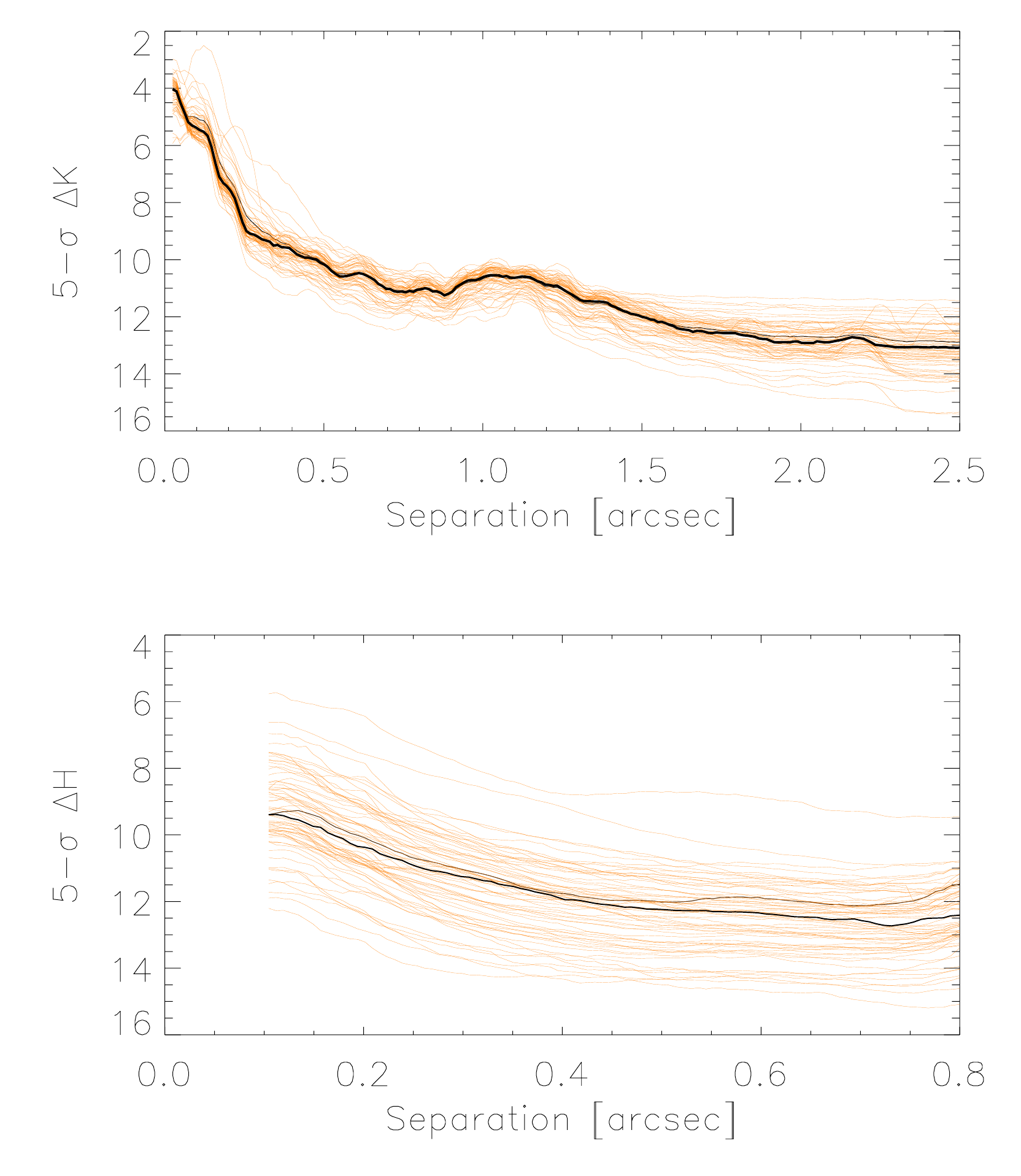}
\caption{Contrast plots obtained for each single target in the sample (thin lines) with IFS at 1.6 $\mu$m (bottom) and IRDIS at 2.15 $\mu$m (top). The thick black line represents the median contrast obtained over the whole sample. Note the different sizes of the X axis. \label{irdiscontrast}}
\end{center}
\end{figure}

The contrast plots were derived on all IRDIS ADI-processed images by measuring the azimutally averaged standard deviation with respect to angular separation. A detection limit of 5-$\sigma$ is considered. Each plot is presented in Fig. \ref{irdiscontrast} (top), and the median of the whole sample per bin of angular separation is displayed in Table \ref{median}. We get a mean magnitude difference of 10 magnitudes at 0.5 arcsec, with a scatter of 1 magnitude along the sample. This corresponds to a contrast range of $\sim2.5\times10^{-3}$ to $\sim4\times10^{-5}$, at 5-$\sigma$.


The IFS contrast was calculated for each object in the sample following the same procedure described in \citet{Mesa2015}. The
contrast plots obtained for each target are shown in Fig.~\ref{irdiscontrast} (bottom) with thin lines. The obtained contrasts  range
from very good ($\sim10^{-6}$ at a separation of 0.5 arcsec) to very poor ($\sim10^{-3}$ at the same separation). The
median contrast calculated over the whole sample is shown as a thick black line in Fig.~\ref{irdiscontrast} (bottom) and Table \ref{median}, with a typical value of $\sim1.25\times10^{-5}$ at 0.5 arcsec.\\

\begin{figure}[t]
\begin{center}
\includegraphics[width=\columnwidth]{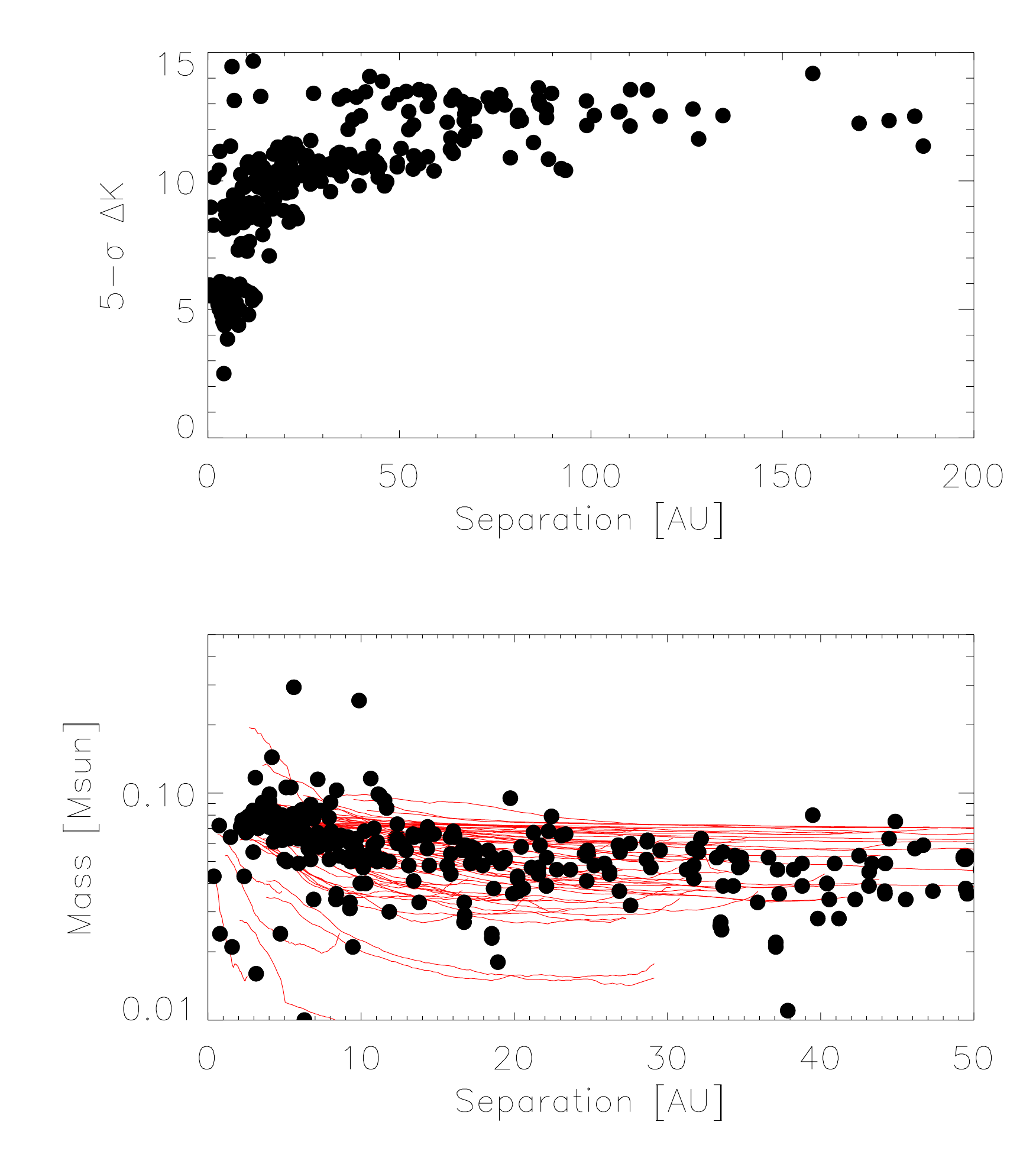}\hfill
\caption{Top: detection limits at 5-$\sigma$ with respect to physical separation, for 0.12-0.24-0.5 and 1" angular separations and all stars included in this study (IRDIS only). Bottom: corresponding minimum mass and physical separation of each star in the sample, derived from IRDIS (black points) and IFS (red lines) detection limits.  \label{mass}}
\end{center}
\end{figure}

This 5-$\sigma$ detection limit is then translated into minimum mass with the aid of BT-Settl stellar models \citep{2014IAUS..299..271A}, and angular separations are converted into physical separations using the system's distance given in Table \ref{tab_sample1} and \ref{tab_sample2}. Figure \ref{mass} summarizes the detection limits for the full sample,  which are also given at a few specific separations in Tables \ref{log} and \ref{logctd}. 

The two stars for which the detection limit is very shallow (Figure \ref{mass} bottom) are 91Aqr and HIP\,105854, two nearby giant stars. Their luminosity limits the magnitude of any companion, even at high contrasts. For all other cases, we get upper limits of masses in the range 0.1-0.15 \msol\, at the inner separations of typically 3 to 8au, and lower than 0.07 \msol\, beyond 10au.

\section{Discussion and perspectives}

We have compared two unbiased samples, one consisting of exoplanet hosts with an eccentric outer planet, and the other made up of exoplanet hosts with circular outer planets. We performed high-contrast imaging on both samples of stars in search for stellar companions, and {\it a posteriori} added the information of known visual binaries to have the most complete census of the systems. We also inferred the detection limits for each star, expressed in the mass and physical distance of any putative bound companion. Observations were made using  SPHERE on the VLT in 2015 and 2016 with both the IRDIS imager and the IFS spectro-imager.\\

Although the program was executed as a bad weather filler, much of the data is of good quality and detection limits are consistently better than 10 to 12 magnitudes at 0.5 arcseconds from the primary. For the data cubes where quality was variable, we applied criteria of frame selection before combining the data. A homogeneous treatment was then performed on the 68 data sets, allowing us to draw self-consistent detection analyses. The detection limits drawn from this study, presented in Fig. \ref{mass} and Tables \ref{log} and \ref{logctd}, show that stellar companions in the range from a few au to 50au would have been detected, except for two stars that are giant  (91 Aqr and HIP\,105854). So, combining the already-known stellar companions and the data of this program, we have a global status on the existence of stellar companions in both our samples. %

In two systems (HD\,7449 and HD\,211847), we detected a close-by stellar companion. For HD\,7449, we find a secondary separation of $\sim 20.4$au in agreement with \citet{2016ApJ...818..106R}. This confirms that this system could be one of the most extreme planet-hosting binaries with a ratio between the companion's and planet's separations of  perhaps less than ten. It may have an architecture  similar to that of a small group of other extreme systems, such as $\gamma$ Cephei, HD196885, or HD41004, with a companion at $\sim20$au and a giant planet around $\sim2$au, and could present a challenge to the canonical planet formation scenario \citep{2014arXiv1406.1357T}. This ratio of semi-major axes should, however, be confirmed by further follow-up of the stellar companion, for which the orbit is not yet fully constrained.

HD\,211847 was first identified from its radial-velocity time series, showing the existence of a long-period eccentric companion of $\sim$ 19 \MJ\ minimum mass \citep{2011A&A...525A..95S}. The source found in direct imaging is at the correct position to match the radial-velocity companion. This implies that this companion at $\sim$11au is a relatively late M dwarf star with an orbit close to face-on. However, looking back into the radial-velocity time series, it is possible that another short-period signal exists here that requires further data and analysis. 
\\

The eccentric sample includes 34 stars (from 35 originally, but HD\,211847 may not be an exoplanet host anymore) and 41 (resp., 40) planets. Ten (29\%) of the host stars are visual binaries at any separation from 20 to 9200au. In contrast, the comparison circular sample includes 27 stars and 43 exoplanets, and only 3 (11\%) visual binaries, at separations ranging from 400 to 1000au. It is striking that more visual binaries are found in the first sample and more exoplanets are present in the second sample. It is not relevant to quantify the significance of this second observation as the exoplanet population in these systems may still be incomplete. Calculating the statistical T-test between both samples for their binary fraction, however, is possible: we get a statistical power of 4.6 while the significance threshold is 1.7; so the null hypothesis of a unique distribution of binaries amongst both samples is excluded. In other words, the proportion of binaries in the eccentric and circular samples are significantly different, despite no other bias than the eccentricity of their outer planet. Also, our detection limits having excluded any other stellar companion in the vicinity of the observed stars, it is striking that the majority of high-eccentricity planets do not orbit stars in multiple systems. The origin of eccentricity in those systems that have a single host star has yet to be explained, possibly involving yet unfound planets. In the case of HD\,4113 (part of our eccentric sample), a long-term trend shows the presence of a massive companion that is yet undetected in direct imaging \citep{2008A&A...480L..33T}.

Fig.\ref{mass} indicates that our survey would have detected any potential companion of mass $\geq 0.1M_{\odot}$ with a separation in the 0.05 to 50au range. Interestingly, this range corresponds to where the distribution of binary separation peaks for solar-type stars \citep{2010ApJS..190....1R}. Considering the separation distribution given in Fig.13 of that paper, we would expect $\sim 25\%$ of all binaries to fall within that range. Thus, given a binarity rate of approximately $50\%$ \citep{2010ApJS..190....1R}, we would expect approximately eight companions in this separation range for the 35+27=62 stars of our combined samples, instead of only one (HD\,7449).
Our results thus confirm the paucity of planet-host binaries with less than $\sim$50-100au separation found in previous studies. \citet{2014ApJ...783....4W,2014ApJ...791..111W} and \citet{2016AJ....152....8K} have reported on such a paucity for a sample of Kepler Objects of Interest, while \citet{eggenberger2007} has also previously reported a less frequent occurrence of binaries with mean semi-major axes between 35 and 100 au around planet-host stars. Finally, \citet{2007A&A...468..721B} also found a lower frequency of planets around binaries with separation smaller than 50-100 au with respect to single stars and members of wide binaries. This first study included a comparison sample of stars for which no radial-velocity companions were detected, while \citet{2014ApJ...783....4W} and \citet{2016AJ....152....8K} based their analysis on a comparison with the binary distribution over the field stars. In our analysis, the comparison is made over the population of known exoplanetary systems with a distinction between dynamically active and inactive systems based on the measured eccentricity of the outer companion. Although our sample is smaller than in those other studies, it is interesting to note that results do not strongly differ and, indeed complement one another. 

In addition, there are 12 stars with at least one identified {candidate} companion in the eccentric sample, and only 6 in the circular sample. This difference is interesting, although most of these candidate companions are very unlikely gravitationally bound to the planetary system, as the dependence with galactic latitude shows. Only HD\,96167 might have a low-mass companion, but its identification requires astrometric follow-up observations for confirmation. HD\,96167, having an eccentric planet (1.3au, eccentricity 0.71 and minimum mass 0.68 \MJ), might represent an interesting dynamical architecture, although the outer companion is seen at a large separation of $\sim$ 290au (if it is a bound system).

\begin{appendix}
\section{Additional technical details and median contrast}
\label{appA}
\subsection{Observation table}
  Tables \ref{log} and \ref{logctd} give the details of the observations presented in this study. Together with the date, mean seeing, and rotation angle obtained for each sequence, the detection limits at various separations are given in the H and K bands.

\begin{table*}
\caption{Log of the SPHERE/IRDIFS observations  and magnitude contrasts observed at a few specific angular separations. }
\label{log}
\begin{tabular}{lccccccccccc}
\hline
Name & MJD-OBS & Seeing &Rot&  $\Delta$H &$\Delta$H &$\Delta$H &$\Delta$K &$\Delta$K &$\Delta$K &$\Delta$K &$\Delta$K \\
       &                 &arcsec  &degrees& 0.15"          & 0.30"         &0.60"         &0.15"       &  0.30"       &  0.60"      &   1.2"        & 2.4"  \\ 
\hline
\hline
GJ 3021 & 57305.10& 1.25  &6.6 & 9.14 &10.87 &12.03&6.98 & 9.50 &10.73 &11.20 &13.53\\
HD 4113 & 57303.18  & 0.85  &21.6  &  12.24 &13.74 &14.10 & 7.02 &10.14 &11.55 &11.83 &13.27\\
HD 7449 & 57304.07 & 1.00  &2.4 &7.58 &10.18 &11.56 &  6.35 &9.06 &9.87 &10.11 &12.06\\
HD 20782 & 57309.20 & 1.23 &65.2 &  8.73 &10.89 &12.19 & 6.65 & 9.10 &10.11 &10.87 &13.15\\
$\epsilon$ Eridani & 57312.19 & 1.56 &11.3 & 11.24 &12.82 &14.30 &  6.77 & 9.50 &10.41 &11.60 &15.36\\
HD 28254 & 57324.29 & 1.22 & 4.5  & 8.97 &11.11 &12.14 & 6.49 & 9.26 &10.13 &10.82 &12.89\\
HD 30562 & 57312.24 & 1.19 & 4.9& 8.40 &10.62 &12.35 & 6.62 & 9.15 &10.44 &11.06 &14.05\\
HD 33283 & 57309.30 & 0.93 & 6.1    &   9.81 &11.30 &12.05 & 6.93 & 9.17 & 9.86 &10.82 &12.79\\
HD 39091 & 57352.22 & 1.33 & 6.6 &  9.21 &11.09 &12.37 & 6.23 & 8.87 &10.43 &10.68 &13.86\\
HD 43197 & 57307.34 & 1.64 & 23.6 & 10.08 &11.87 &12.53 & 6.60 &9.23 & 10.34 &12.01 &12.54\\
HD 44219 & 57310.30 & 0.95 & 3.7  &  9.98 &12.21 &13.36 & 6.81 & 9.03 &10.62 &10.83 &13.40\\
HD 66428 & 57347.30 & 0.76 & 7.5 &  10.04 &11.97 &12.89 & 5.99 & 8.90 &10.72 &11.38 &12.79\\
HD 86226 & 57375.33 & 0.84 & 0.1&8.36 &10.47 &11.18 & 6.54 & 9.35 &10.41 &10.38 &12.70\\
HD 86264 & 57122.06 & 1.18 & 19.8& 10.64 &12.66 &13.29 & 6.75 & 9.41 &11.53 &12.08 &12.59\\
HD 96167 & 57122.09 & 1.65 & 12.2 &  10.77 &12.53 &13.04 & 6.26 & 9.17 &11.09 &11.72 &12.26\\
HD 98649 & 57182.96 & 1.88 & 3.8  &  9.20 &10.28 &11.94 & 6.91 & 9.20 &10.27 &10.96 &12.69\\
HD 106515 & 57130.11 & 1.87 & 7.8 &  8.95 &11.13 &12.27 & 6.72 & 9.33 &10.76 &11.45 &13.06\\
HD 108147 & 57220.99 & 0.71 & 4.9 &  10.04 &12.65 &13.53 & 6.60 & 9.56 &10.69 &10.92 &13.36\\
HD 129445 & 57129.26 & 1.41 & 7.34&8.22 &10.39 &11.45 & 5.58 & 7.86 &10.69 &11.04 &11.80\\
HD 142022 & 57197.16 & 0.82 & 2.4  &  8.50 &10.17 &11.30 & 6.28 & 9.37 &10.55 &10.94 &13.05\\
HD 142415 & 57197.20 & 0.83 &  4.1&9.63 &11.42 &12.23 & 6.70 & 9.29 &10.47 &10.99 &13.13\\
HD 148156 & 57134.37 & 0.90 & 4.94 &  10.34 &12.03 &12.92 & 6.58 & 9.96 &11.33 &11.34 &12.16\\
HD 154672 & 57134.39 & 0.94 & 5.2 &  10.22 &11.88 &12.72 & 7.05 &10.09 &11.19 &11.37 &12.85\\
HD 157172 & 57309.01 & 0.78 & 0.3  &  7.85 &10.05 &11.29 & 6.80 & 9.32 &10.15 &10.57 &12.46\\
HD 168443 & 57124.40 & 0.98 & 15.37 &  12.68 &14.20 &14.41 & 7.75 &10.62 &12.01 &12.25 &13.29\\
HD 181433 & 57309.03 & 0.70 & 7.36 &  9.53 &11.48 &12.34 & 6.46 & 9.03 &10.04 &10.75 &12.96\\
HD 187085 & 57306.01 & 1.24 & 34.4 &  11.46 &13.32 &14.11 & 6.68 & 9.76 &11.07 &11.63 &13.93\\
HD 196067 & 57191.40 & 1.38 & 3.4  &  8.92 &10.36 &12.10 & 5.86 & 8.69 &10.15 &10.89 &13.24\\
HD 210277 & 57299.10 & 0.76 & 14.5 &  11.46 &13.77 &14.19 &6.74 & 9.43 &11.35 &11.80 &14.29\\
HD 211847 & 57183.39 & 1.31 & 6.7 & 10.37 &11.07 &12.87 & 6.69 & 8.34 &10.50 &11.08 &12.21\\
HD 215497 & 57304.05 & 1.06 & 9.09&9.82 &11.65 &12.46 &6.71 & 9.40 &10.22 &11.12 &13.22\\
HD 217107 & 57183.37 & 1.55 & 3.2 &  10.48 &11.14 &12.92 & 6.94 & 9.26 &10.22 &10.68 &13.90\\
HD 219077 & 57182.39 & 1.29 & 3.3 & 10.37 &11.27 &12.80 &6.89 & 9.41 &10.22 &10.87 &14.08\\
HD 222582 & 57193.41 & 1.07 & 5.3 & 9.40 &10.9 &11.57 &  4.60 & 7.90 & 9.87 &10.75 &12.48\\
\hline
\end{tabular}
\end{table*}

\begin{table*}
\caption{Log of the SPHERE/IRDIFS observations and magnitude contrasts observed at a few specific angular separations (cont'd). }
\label{logctd}
\begin{tabular}{lccccccccccc}
\hline
Name & MJD-OBS & Seeing &Rot&  $\Delta$H &$\Delta$H &$\Delta$H &$\Delta$K &$\Delta$K &$\Delta$K &$\Delta$K &$\Delta$K \\
       &                 &arcsec  &degrees& 0.15"          & 0.30"         &0.60"         &0.15"       &  0.30"       &  0.60"      &   1.2"        & 2.4"  \\ 
\hline
\hline
HD 1461 & 57299.13 & 0.73 & 10.8  & 10.19 &12.65 &13.38 & 6.13 & 9.05 &10.88 &11.73 &14.16\\
HD 4208 & 57308.08 & 0.82 & 2.3& 9.27 &10.85 &12.39 & 6.77 & 9.71 &10.64 &10.76 &13.47\\
HD 10180 & 57300.09 & 0.83 & 8.0 &  10.19 &12.83 &13.48 &  6.32 & 9.34 &10.41 &11.07 &13.66\\
HD 11964 & 57306.12 & 1.05 & 2.13& 8.68 &10.32 &11.91 & 6.89 & 9.17 &10.63 &10.72 &13.45\\
HD 20794 & 57309.22 & 1.43 & 22.8 & 11.93 &13.34 &14.70 & 6.55 & 9.16 &10.88 &12.43 &15.37\\
HD 23079 & 57312.17 & 1.48 & 7.8  &  9.93 &11.62 &12.34 & 6.88 & 9.44 &10.26 &11.07 &13.41\\
HD 38801 & 57312.26 & 1.27 & 50.0&9.90 &11.37 &12.06 & 6.69 & 9.43 &10.36 &11.03 &12.47\\
BD-061339 & 57309.36 & 1.16 &6.1&9.40 &11.09 &11.91 &5.81 & 8.92 & 9.94 &10.93 &12.73\\
HD 47186 & 57343.29 & 1.17 & 5.5 &  10.13 &11.87 &12.75 & 6.28 & 9.37 &10.87 &10.83 &13.04\\
HD 60532 & 57343.30 & 1.40 & 2.2&7.28 & 9.96 &11.60 & 6.31 & 9.15 &10.46 &10.50 &14.03\\
HD 65216 & 57350.34 & 1.01 & 8.9&9.74 &11.26 &11.97 & 5.29 & 8.37 &10.08 &10.73 &12.01\\
HD 73256 & 57378.36 & 0.77 & 53.78 & 10.93 &12.72 &13.26 & 6.63 & 9.40 &10.90 &11.36 &13.31\\
HD 75289 & 57395.17 & 0.67 & 2.8 &  10.23 &12.28 &13.63 & 6.23 & 8.75 &10.47 &10.48 &13.39\\
HD 76700 & 57347.35 & 0.90 & 2.9&8.12 &10.13 &11.36 & 6.28 & 8.89 &10.58 &10.67 &12.60\\
HD 82943 & 57122.04 & 1.21 & 16.7&  11.77 &13.43 &13.61 & 7.27 & 9.45 &11.31 &11.68 &12.26\\
HD 83443 & 57377.25 & 0.40 & 2.49&10.17 &11.32 &12.49 & 6.50 & 9.30 &10.66 &10.24 &12.58\\
HD 85390 & 57343.33 & 2.10 & 7.2&6.98 & 8.49 &10.49 & 4.99 & 8.19 & 9.95 &10.62 &11.93\\
HD 86081 & 57347.33 & 0.78 & 1.2  &  8.24 & 9.96 &10.88 & 6.42 & 9.17 &10.46 &10.37 &11.43\\
HD 92788 & 57183.98 & 1.80 & 2.5&7.90 & 9.60 &11.24 & 6.61 & 9.25 & 9.82 &10.53 &13.18\\
HD 104067 & 57138.07 & 1.84 & 5.4& 10.38 &11.84 &12.96 & 6.68 & 9.88 &10.88 &11.34 &13.57\\
HD 109749 & 57221.02 & 0.66 & 1.9 &  8.05 &10.57 &11.42 & 6.84 & 9.49 &10.57 &10.79 &11.56\\
HD 117618 & 57197.18 & 0.78 &2.0  & 9.20 &10.78 &11.83 & 6.90 & 9.40 &10.49 &11.01 &13.17\\
HD 121504 & 57270.99& 0.72 & 3.85&  9.09 &10.62 &11.53 & 6.20 & 9.18 &10.39 &10.79 &12.97\\
HD 134060 & 57140.39 & 0.90 & 2.6&9.46 &11.62 &12.34 & 6.07 & 9.69 &10.75 &11.20 &13.15\\
HD 150433 & 57135.36 & 1.06 & 3.8&8.99 &11.22 &12.45 &  6.62 &10.15 &10.83 &11.05 &13.53\\
HD 159868 & 57133.39 & 1.05 & 5.4  &  9.68 &11.17 &12.71 & 5.91 & 8.92 &10.76 &11.18 &13.71\\
HD 192263 & 57301.05 & 1.08 & 7.5  &  9.98 &12.00 &13.22 & 6.89 & 9.50 & 9.94 &10.77 &12.64\\
HIP 105854 & 57182.37 & 1.30 & 8.9  &  11.56 &12.93 &14.28 &6.87 & 9.51 &10.42 &11.34 &14.61\\
HD 212301  &  57182.41& -& 2.56&8.18  &  9.59 &10.99 &6.99 & 9.72 &10.87 &11.01 &11.86\\
91 Aqr&  57183.41 & 1.39 & 8.8&  9.80 &  11.17 &11.90 &  6.75 & 9.05 &10.19 &10.55 &14.28\\
\hline
\end{tabular}
\end{table*}

\subsection{Frame selection}

\begin{figure*}
\begin{center}
\includegraphics[width=1.3\columnwidth]{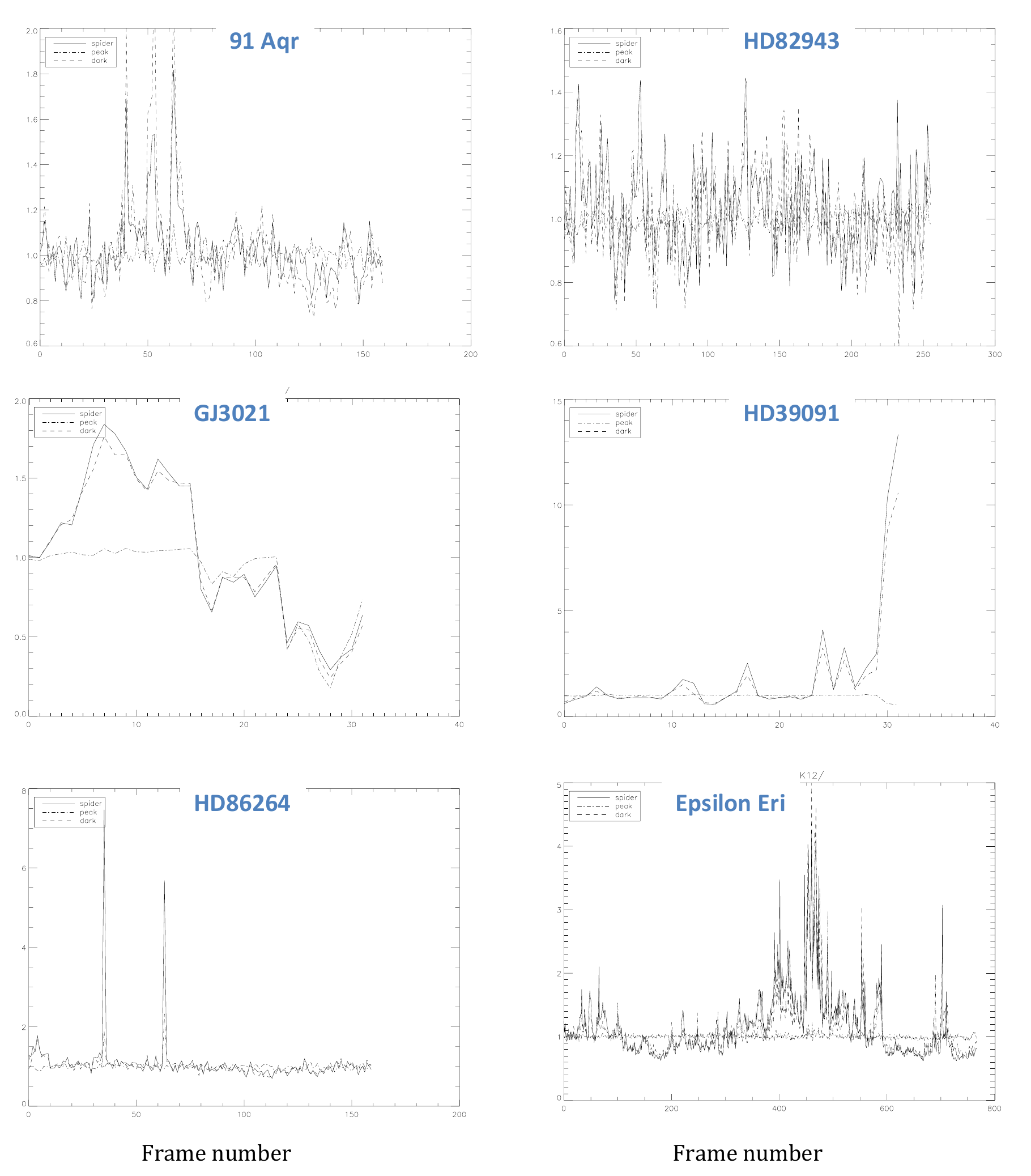}\hfill
\caption{Examples of flux criteria in various places of the image along the IRDIS data cubes used for the selection of valid frames in the combination processing (see text). The plain line shows the flux in the image close to the spider (indicating a pupil misalignment), the dashed line shows the flux in the darkest region of the image, and the dot-dash line shows the peak flux in the center of the target. \label{selectionframe}}
\end{center}
\end{figure*}

We used the following method to select the frames: \begin{itemize}
    \item First, we estimated the difference image between each frame and the cube median; referred to as the  residual frame in the following.
    \item We measure three quantities in the residual frames: i) the maximum of the peak in the central area, ii) the flux at the location of the residual diffraction of the spiders, and iii) the average flux in the darkest, AO-corrected part of the image, above and below the peak. Each quantity is normalized by the median of the time sequence, for each cube.
    \item We plotted the time evolution of these quantities for each cube and determined threshold for each, beyond which the data should be removed. This trade-off analysis aims at keeping the maximum number of frames while not increasing the noise in the combined image. It is usually quite clear from the data where the line should be drawn, as illustrated in some of the examples shown in Fig. \ref{selectionframe}. 
    \item We removed all frames when either the normalized peak was lower than 0.5, the normalized spider flux was more than 2.0, or the normalized dark flux was more than 2.0. These values turned out to be a suitable compromise for all data cubes. 
\end{itemize}

The number of removed frames for each observation is given in Tables \ref{log} and \ref{logctd} for reference. Not many sequences are affected by frame selection, an indication that this filler program was used in reasonably good conditions in most cases. For some data cubes, however, the frame selection removed more than half of the data.

\subsection{Median detection limit}
 The contrast plots as a function of separation obtained in this analysis are very similar for all objects, with offsets due to the data quality and/or the star magnitude. Table \ref{median} gives the median contrast plot in H and K magnitudes, obtained respectively from IFS-YH and IRDIS-K12 configurations. These values correspond to the thick black curves shown on Figure \ref{irdiscontrast}.

\begin{table*}
\caption{Median contrast curve as a function of angular separation in YH from IFS data, and in K from IRDIS data. These curves are calculated from the 64 data sets analyzed in this article, and shown on Figure  \ref{irdiscontrast}.}
\label{median}
\begin{tabular}{lcc}
\hline
Sep & $\Delta$H & $\Delta$K \\
arcsec& mag & mag \\
\hline
    0.10 &   9.39 &   5.39\\
    0.15 &   9.75 &   6.21\\
    0.20 &  10.37 &   7.50\\
    0.25 &  10.91 &   8.84\\
    0.30 &  11.26 &   9.25\\
    0.35 &  11.55 &   9.49\\
    0.40 &  11.92 &   9.67\\
    0.45 &  12.11 &   9.93\\
    0.50 &  12.24 &  10.17\\
    0.55 &  12.30 &  10.59\\
    0.60 &  12.35 &  10.49\\
    0.65 &  12.47 &  10.65\\
    0.70 &  12.57 &  11.02\\
    0.75 &  12.64 &  11.13\\
    0.80 &  12.41 &  11.05\\
    0.90 &  - &  11.15\\
    1.00 &  - &  10.62\\
    1.10 &  - &  10.63\\
    1.20 &  - &  10.88\\
    1.30 &  - &  11.36\\
    1.40 &  -  & 11.57\\
    1.50 &  -  & 12.00\\
    1.60 &  -  & 12.33\\
    1.70 &  -  & 12.54\\
    1.80 &  -  & 12.59\\
    1.90 &  -  & 12.82\\
    2.00 &   - &  12.91\\
    2.10 &   - &  12.87\\
    2.20 &   - &  12.78\\
    2.30 &   - &  13.07\\
    2.40 &   - &  13.07\\
    2.50 &   - &  13.08\\
\hline
\end{tabular}
\end{table*}

\end{appendix}

\begin{acknowledgement}
The authors thank the SPHERE  team for an efficient and reliable instrument, and the ESO science operation support for their help with the scheduling and execution of observations.  We are grateful to the referee for her/his constructive comments that greatly improved the article. DM and SD acknowledge support from the "Progetti Premiali" funding scheme of the Italian Ministry of Education, University, and Research.
\end{acknowledgement}

\bibliographystyle{aa} 
\bibliography{moutou_references}

\end{document}